**Magnetic and Orbital Orders Coupled to Negative Thermal Expansion in Mott**

**Insulators, $Ca_2Ru_{1-x}M_xO_4$ (M = Mn and Fe)**

T. F. Qi[1,2], O. B. Korneta[1,2], S. Parkin[1,3], Jiangping Hu[4] and G. Cao[1,2*]

[1] Center for Advanced Materials, University of Kentucky

[2] Department of Physics and Astronomy, University of Kentucky

[3] Department of Chemistry, University of Kentucky

[4] Department of Physics, Purdue University

$Ca_2RuO_4$ is a structurally-driven Mott insulator with a metal-insulator transition at $T_{MI} = 357K$, followed by a well-separated antiferromagnetic order at $T_N = 110$ K. Slightly substituting Ru with a 3d transition metal ion M effectively shifts $T_{MI}$ by weakening the orthorhombic distortion and induces either metamagnetism or magnetization reversal below $T_N$. Moreover, M doping for Ru produces negative thermal expansion in $Ca_2Ru_{1-x}M_xO_4$ (M = Cr, Mn, Fe or Cu); the lattice volume expands on cooling with a total volume expansion ratio, $\Delta V/V$, reaching as high as 1%. The onset of the negative thermal expansion closely tracks $T_{MI}$ and $T_N$, sharply contrasting classic negative thermal expansion that shows no relevance to electronic properties. In addition, the observed negative thermal expansion occurs near room temperature and extends over a wide temperature interval up to 300 K. These findings underscores new physics driven by a complex interplay between orbital, spin and lattice degrees of freedom.

**PACs**: 71.30.+h; 75.47.Lx; 65.40.De

## I. Introduction

The Coulomb interaction U is generally comparable to the 4d-bandwidth W in the 4d-based ruthenates, $A_{n+1}Ru_nO_{3n+1}$ with n = 1,2,3,$\infty$ (A = Ca or Sr), which leaves them precariously balanced on the border between metallic and insulating behavior, and/or on the verge of long-range magnetic order. A common characteristic of these materials is that underlying physical properties are critically linked to the lattice and orbital degrees of freedom, and tend to exhibit a non-linear or giant response to modest lattice changes. This is illustrated by observations of a *p*-wave superconducting state in $Sr_2RuO_4$ [1] and a first-order metal-insulator (MI) transition at $T_{MI}$ = 357 K which is followed by a well separated antiferromagnetic transition at $T_N$ = 110 K in $Ca_2RuO_4$ that is more structurally distorted due to the smaller ionic radius $r_{Ca} < r_{Sr}$ [2, 3].

In $Ca_2RuO_4$, the MI transition induces a radical change in electrical resistivity $\rho(T)$ by over nine orders of magnitude below $T_{MI}$, which also marks a concomitant and violent structural transition between a high-T tetragonal and a low-T orthorhombic phase at $T_{MI}$, or a strong cooperative Jahn-Teller distortion [2-4]. Below $T_{MI}$, the **a**-axis contracts by 1.5%, but the **b**-axis expands by 3% on cooling, over an temperature interval of 250 K; the combined effect of these conflicting uniaxial thermal expansions is to drive an increasingly strong orthorhombic distortion in the basal plane that shatters single-crystal samples and contracts the lattice volume by 1.3% as temperature is lowered from 400 K to 70 K [2, 3]. It is already established that such a strong cooperative Jahn-Teller distortion removes the degeneracy of the three Ru $t_{2g}$ orbitals ($d_{xy}$, $d_{yz}$, $d_{zx}$) via a transition to orbital order that, in turn, drives the MI transition at $T_{MI}$ = 357 K [5-9], which is followed by an antiferromagnetic order at $T_N$ = 110 K that is well below $T_{MI}$ = 357 K, sharply contrasting classic Mott insulators that undergo simultaneous transitions to antiferromagnetic order and an insulating state at $T_{MI}$. This behavior, which is also observed in its sister compound, $Ca_3Ru_2O_7$ [10], separates the layered $Ca_{n+1}Ru_nO_{3n+1}$ (n = 1 and 2) from classic Mott insulators as a unique archetype of Mott insulators in which the MI transition is not primarily driven by antiferromagnetic order. Controversy over the exact nature of the orbital state remains, but the extraordinary sensitivity of $T_{MI}$ to modest changes in lattice parameters or modest pressure [5-9, 11-13] clearly indicate that the lattice and orbital degrees of freedom play a decisive role in the new physics that drives $Ca_2RuO_4$.

Indeed, we have recently observed that modest Cr doping for Ru in $Ca_2RuO_4$ not only changes $T_{MI}$ and the magnetic behavior below $T_N$ but also triggers off a two-step, abrupt negative thermal expansion (NTE) at $T_{MI}$ and $T_N$ and a nearly zero thermal expansion between $T_C$ and $T_{MI}$, giving rise to a total volume expansion ratio $\Delta V/V \approx 1$ % on cooling over $90 < T < 220$ K. This behavior indicates that the observed NTE is strongly coupled to the orbital and magnetic orders [14], in contrast to classic NTE primarily driven by phonon modes or geometry effects [15-21]. It is noted that the origin of the NTE recently observed in $BiNiO_3$ is attributed to charge transfer [22].

In this paper, we report results of our more recent study that we have extended to include other 3d ion dopants. The central findings of this study are that the strong coupling of the NTE to orbital and magnetic orders is in fact a common occurrence in a class of materials, $Ca_2Ru_{1-x}M_xO_4$ where M = Mn, Fe, or Cu, and that M doping effectively changes $T_{MI}$ by weakening the orthorhombic distortion, and induces unusual magnetic behavior such as metamagneticsm and magnetization reversal below $T_N$. In the following, we present and discuss the underlying physical properties and the NTE observed in Mn and Fe doped $Ca_2RuO_4$. The simultaneous occurrence of the NTE and orbital and magnetic orders underscores a mechanism driven by complex electronic correlations; this mechanism is qualitatively discussed.

## II. Experimental Techniques and Structural Refinements

The single crystals of $Ca_2Ru_{1-x}M_xO_4$ with M = Mn and Fe and $0 \leq x \leq 0.25$ were grown using a floating-zone optical furnace; details of single-crystal growth are described elsewhere [14, 23]. Our single-crystal x-ray diffraction study of $Ca_2Ru_{1-x}M_xO_4$ was performed as a function of temperature between 90K and 430K using a Nonius-Kappa CCD single-crystal X-ray diffractometer. The structures were refined by the SHELX-97 programs [24, 25]. All structures affected by absorption and extinction were corrected by comparison of symmetry-equivalent reflections using the program SADABS [25]. It needs to be emphasized that the single crystals are of high quality and there is no indication of any mixed phases or inhomogeneity in all doped single crystals studied. The presence of any mixed phases or inhomogeneity in the single crystals would not allow any converging structural refinements. Some of the experimental and refinement details for three different dopants at a similar doping concentration are given in **Table 1.** (The data Cr doping is also listed for the purpose of comparison.) Single crystal diffraction data collected from $Ca_2Ru_{1-x}M_xO_4$ samples are indexed on the basis of the orthorhombic symmetry shown in **Table 1**. Note that errors of the last digit given in parentheses represent only statistical errors, not systematic errors, which may arise from correlations between parameters. The rest of the structure data including x and temperature dependence will be published elsewhere [26]. Error bars for the structural data presented here are smaller than the symbols, therefore are not included. Chemical compositions were determined by Energy Dispersive X-ray analysis (EDX); the EDX data confirmed

**Table 1** Experimental and refinement details of $Ca_2Ru_{1-x}M_xO_4$ (M=Cr, Mn and Fe)

| | $Ca_2Ru_{1-x}\mathbf{Cr}_xO_4$ | $Ca_2Ru_{1-x}\mathbf{Mn}_xO_4$ | $Ca_2Ru_{1-x}\mathbf{Fe}_xO_4$ |
|---|---|---|---|
| **Crystal data** | | | |
| Chemical formula | $Ca_4Ru_{1.87}$ $Cr_{0.13}O_8$ | $Ca_4Ru_{1.9}$ $Mn_{0.1}O_8$ | $Ca_4Ru_{1.88}$ $Fe_{0.12}O_8$ |
| $M_r$ | 474.02 | 490.46 | 490.46 |
| Crystal system, space group | Orthorhombic, *Pbca* | Orthorhombic, *Pbca* | Orthorhombic, *Pbca* |
| Temperature (K) | 90K | 90K | 90K |
| $a, b, c$ (Å) | 5.3917 (2), 5.5157 (2), 11.8804 (4) | 5.3856 (1), 5.5875 (2), 11.7772 (4) | 5.4045 (1), 5.5809 (2), 11.7718 (6) |
| $\alpha, \beta, \gamma$ (°) | 90, 90, 90 | 90, 90, 90 | 90, 90, 90 |
| V (Å$^3$) | 353.31 (2) | 354.40 (2) | 355.06 (2) |
| Z | 2 | 2 | 2 |
| Radiation type | Mo *Kα* | Mo *Kα* | Mo *Kα* |
| $\mu$ (mm$^{-1}$) | 6.99 | 7.17 | 7.15 |
| Crystal size (mm) | 0.05 ×0.05 ×0.05 | 0.01 × 0.05 × 0.05 | 0.01 × 0.1 × 0.03 |
| **Data collection** | | | |
| Diffractometer | Nonius KappaCCD diffractometer | Nonius KappaCCD diffractometer | Nonius KappaCCD diffractometer |
| Absorption correction | Multi-scan SADABS(Sheldrick, 1996) | Multi-scan SADABS(Sheldrick, 1996) | Multi-scan SADABS(Sheldrick, 1996) |
| No. of measured, independent and observed[$I > 2\sigma(I)$] reflections | 7241, 398, 352 | 4888, 350, 294 | 4924, 398, 289 |
| $R_{int}$ | 0.030 | 0.055 | 0.046 |
| **Refinement** | | | |
| R[$F^2 > 2\sigma(F^2)$], $wR(F^2)$, $S$ | 0.016, 0.046, 1.13 | 0.067, 0.173, 1.30 | 0.032, 0.089, 1.14 |
| No. of reflections | 398 | 350 | 398 |
| No. of parameters | 36 | 35 | 36 |
| No. of restraints | 0 | 0 | 0 |
| $\Delta\rho_{max}, \Delta\rho_{min}$ (e Å$^{-3}$) | 0.66, -0.60 | 2.99, −2.90 | 2.05, −1.35 |

The high homogeneity of all crystals studied. Measurements of magnetization M(T,H), heat capacity C(T) and electrical resistivity ρ(T) for $1.7 < T < 400$ K were performed using either a Quantum Design Physical Property Measurement System or Magnetic Property Measurement System.

### III. Results and Discussions

### A. Ca$_2$Ru$_{1-x}$Mn$_x$O$_4$ with $0 < x \leq 0.25$

**Fig. 1a** exhibits the lattice parameters for the **a**-, **b**- and **c**-axis at T = 90 K as a function of Mn concentration x ranging from 0 to 0.25 for single-crystal Ca$_2$Ru$_{1-x}$Mn$_x$O$_4$. Like Cr doping **[14, 23]**, Mn doping preserves the low temperature orthorhombic symmetry (*Pbca*) but weakens the orthorhombic distortion by reducing the difference between the **a**- and **b**-axis or [**b**–**a**] from 0.25 Å for x = 0 to 0.07 Å for x = 0.25. Furthermore, the unit cell volume V for x $\geq$ 0.085 is larger at 90 K than at 295 K, indication NTE; but the NTE diminishes when x approaches 0.25, as shown in **Figs. 1b** and **1c.**

We now focus on the coupling between the NTE and MI transition at T$_{MI}$ and the magnetic order at T$_N$ by examining two sets of representative data for x = 0.10 and 0.25. At x=0.10, the NTE occurs along both the **a**-and **b**-axis; and this combined effect results in an overall volume expansion ratio ΔV/V ≈ 0.8% on cooling with an onset in close proximity to the MI transition that occurs at T$_{MI}$ = 380 K (ρ$_{ab}$ is in log scale), as shown in **Fig. 2**. The coupling of the NTE to the MI transition and the magnetic order is obvious in that V rapidly expands below T$_{MI}$ = 380 K, and exhibits a weak yet well-defined anomaly near T$_N$ = 130 K, where the weak ferromagnetic behavior takes place, as marked by the shaded area and the dashed line in **Fig. 2**. On the other hand, at x = 0.25, the orthorhombicity is considerably weakened and reduced to 225 K, and the MI transition becomes so broadened that it cannot be well defined. Concomitantly, the NTE diminishes with nearly zero thermal expansion or $\Delta V/V < 0.1\%$ when temperature is lowered from 410 K to 90 K, as illustrated in **Fig. 3**. The simultaneous disappearance of both $T_{MI}$ and the NTE reinforces that the NTE and the orbital order are indeed strongly coupled.

**Fig. 4** provides a comparison of the physical properties between several representative Mn concentrations. It is clear that the MI transition increases from $T_{MI} = 357$ K for x = 0 to ~ 380 K for x = 0.085 and 0.10, but the sharpness of the MI transition diminishes for x = 0.14 and vanishes for x = 0.25 (see **Fig. 4a**). Since the MI transition is primarily due to the structural phase transition between the high-T tetragonal to the low-T orthorhombic distortion, the broadening of the MI transition is more likely due to the diminishing of the structural phase transition as Mn doping readily reduces the difference between the **a**- and **b**-axis evident in **Fig.1**. It needs to be pointed out that the MI transition in $Ca_2Ru_{1-x}Mn_xO_4$ is the second-order transition; this is in contrast to the first-order transition that characterizes $Ca_2RuO_4$ **[2-3]** and $Ca_2Ru_{1-x}Cr_xO_4$ **[14, 23]**. The second-order phase transition at $T_{MI}$ for various x is also confirmed by sharp peaks in specific heat C(T), which are narrowly separated, as shown in **Fig.4b**. On the other hand, the magnetic ordering temperature $T_N$ systematically rises with increasing x from $T_N = 120$ K for x=0.085 to 130 K for x=0.10 and eventually 150 K for x = 0.25, as illustrated in the inset in **Fig.4b** and **Fig.4c**. (The systematic change of $T_N$ with x further confirms the homogeneity of the samples studied.) Interestingly, Mn doping induces a metamagnetic transition at $H_C$ and a sizable order moment $\mu_o$; with increasing x, $H_C$ decreases from 5 T for x = 0.085 to 2 T for x = 0.25 whereas $\mu_o$ eventually amounts up to 0.2 $\mu_B$/f.u for x = 0.25, as shown in **Fig.4d.** The metamagnetism indicates a spin canting existent in an antiferromagnetic background, which explains the weak ferromagnetic behavior seen in **Fig. 4c**.

**B. $Ca_2Ru_{1-x}Fe_xO_4$ with $0 < x \leq 0.22$**

Substituting Fe for Ru in $Ca_2RuO_4$ also effectively weakens the orthorhombic distortion, and induces the NTE. **Fig. 5** registers the lattice parameters for $0.08 \leq x \leq 0.22$ taken at 90 K and 295 K. Similar to that for Cr **[14, 23]** and Mn doping (Figs.1 and 3), the effect of the NTE nearly vanishes at x = 0.22 where the orthorhombic distortion is considerably reduced and $T_{MI}$ is no longer well defined.

More detailed data for x = 0.08 as a function of temperature collected over $1.7 < T < 450$ K exhibit the strong NTE along **b**-axis but a much weaker temperature dependence of the **a**-axis (see **Fig. 6a**), and an overall thermal expansion with $\Delta V/V \sim 0.8\%$ on cooling over $90 < T < 390$ K (see **Fig. 6b**). It is now no longer a surprise that the onset of the NTE occurs simultaneously with the MI transition at $T_{MI}$ that is characterized by a strong anomaly in the specific heat C(T) (**Fig.6b**) and $\rho_{ab}$(T) (**Fig.6c**, right scale) at $T_{MI} = 380$ K. It is noted that the NTE peaks near 150 K below which V starts to contract on cooling (**Fig. 6b**), and there is no obvious lattice anomaly near the magnetic order at $T_N = 120$ K, in contrast to the behavior seen in Cr and Mn doped $Ca_2RuO_4$. This difference may be associated with unusual magnetic properties absent in Cr and Mn doped $Ca_2RuO_4$. As seen in **Fig.6c**, $\chi_{ab}$ for x=0.08 shows a peak at $T_N$ seldom seen in other antiferromagnets; this behavior is then evolved into "diamagnetic" behavior for x = 0.12 below $T_N = 120$ K, indicating a rare magnetization reversal (Fig.6c) ($\chi_{ab}$ was measured using a field-cooled sequence; the magnetization reversal exists in $0.08 < x <$ 0.22). A magnetization reversal is highly unusual but is not without precedent; it has been observed in a few ferromagnetic spinels and $Sr_3Ir_2O_7$ [27]. It is conceivable that increasing Fe doping for Ru may lead to two inequivalent magnetic sublattices that are antiferromagnetically coupled; the magnetization reversal could be a result of different temperature dependences of the two individual magnetic sublattices. Without apparent spin canting, the magnetoelastic effect may not be strong enough to causes any additional lattice anomaly that features Cr and Mn doped $Ca_2RuO_4$.

### C. General Trends in $Ca_2Ru_{1-x}M_xO_4$

Substituting Ru with a 3d ion always induces a modest and yet critical NTE along the **a**-axis, as shown in **Fig. 7a**. It is this critical change that leads to a rather sizable thermal expansion ratio $\Delta V/V$ on cooling in $Ca_2Ru_{1-x}M_xO_4$. The fact that this phenomenon does not occur for x = 0 despite the strong effect of the NTE along the **b**-axis underscores how critically M doping "unlocks" strongly buckled $Ru/MO_6$ octahedra and changes the $t_{2g}$ orbital configuration in the basal plane. Indeed, the basal plane Ru/M-O1-Ru/M bond angle $\boldsymbol{\theta}$ drastically decreases below $T_{MI}$ that in turn simultaneously prompts an expansion of the Ru/M-O1 bond length $\boldsymbol{d}$ and the Ru/M-Ru/M distance on cooling; as shown in **Figs. 7b** and **7c** for a representative compound $Ca_2Ru_{1-x}Fe_xO_4$ with x=0.08, respectively. The expansion of $\boldsymbol{d}$ clearly outweighs positive thermal expansion due to longitudinal vibrational modes, allowing ***both*** the **a**- and **b**-axis to expand with cooling while preserving the structural symmetry; and as a result, V rapidly expands on cooling near $T_{MI}$ where the $t_{2g}$ orbital order takes place. Both the NTE and the orbital order in $Ca_2Ru_{1-x}M_xO_4$ closely track the changing orthorhombicity as x changes, and disappear when the orthorhombicity vanishes near $x_c$, a critical doping concentration ($x_c$

= 0.14, 0.25, 0.22 and 0.20 for Cr, Mn, Fe, and Cu, respectively). As for the magnetic state and the spin-lattice effect, it is recognized that an increase in the Ru-O2 bond length along the **c**-axis destabilizes the collinear antiferronagnetic state **[28],** resulting in strongly competing antiferromagnetic and ferromagnetic exchange interactions or spin canting below $T_N$ in doped $Ca_2RuO_4$. Except for Fe doping which does not seem to cause spin canting, the spin-lattice coupling or magnetoelastic effect is strong enough to generate an additional lattice anomaly near $T_N$. It is also noted that the magnitude of the NTE decreases as the atomic number of M increases. This interesting trend may be associated with the fact that with increasing nuclear charge the 3d-orbitals become more contracted, and the 3d-band progressively fills and downshifts away from the Fermi energy $E_F$, thus weakening the overlapping with 4d-band that stays near $E_F$.

It is compelling to attribute the observed NTE to a mechanism where electronic correlations play a critical role. Such a mechanism can be qualitatively discussed as follows. In a Mott insulator, the occurrence of an orbital or magnetic order is always accompanied by electron localization. The electron localization costs kinetic energy of electrons whereas lattice expansion reduces the kinetic energy. Meanwhile, the lattice expansion costs the energy for electron-lattice interaction. NTE could happen if the energy gain from the electron-electron interaction and the lattice expansion can overcome the energy cost from the electron-lattice interaction and the electron localization. When the orbital and/or magnetic order takes place, the energy gain of electrons can be described in terms of the short range coupling parameters between orbital or spin orders; namely, if we use a local exchange effective model to describe the orbital or magnetic order,

$$H = \sum_{<ij>} J_{ij} A_i A_j$$

where $A_j$ represents local spin or orbital moments. The effective coupling parameters $J_{ij}$ are generally determined by virtual hopping processes. Therefore, if the lattice expansion increases $J_{ij}$, the orbital and/or magnetic order can make the NTE more energetically favorable; this is more likely in a multi-orbital system where the Coulomb repulsion U is relatively small for the reasons: **(1)** virtual hopping becomes much more complicated in effective bands due to the mixture of different orbitals; and **(2)** spin-orbit coupling and crystal field effects, which are strongly affected by the lattice expansion, and become comparable to U. $Ca_2Ru_{1-x}M_xO_4$ are multi-orbital systems with comparable U and spin orbit interaction; therefore, the NTE happens, but in a fashion fundamentally different from that of classic NTE materials, as shown in **Figs. 1-7**. A computational result on the values of $J_{ij}$ following the above analysis will be reported elsewhere **[29]**. This work presents convincing evidence that the strong coupling of the NTE to the underlying physical properties exists a class of the Mott insulators, doped $Ca_2RuO_4$, highlighting the new physics yet to be fully understood.

This work was supported by NSF through grants DMR-0856234 and EPS-0814194.

*Correspondence author: cao@pa.uky.edu

**Figure Captions:**

**Fig.1.** For $Ca_2Ru_{1-x}Mn_xO_4$ with $0 \leq x \leq 0.25$, x dependence of **(a)** the lattice parameter **a**-**b**- and **c**-axis (right scale) at T = 90 K, **(b)** unit cell volume V, and **(c)** thermal expansion ratio $\Delta V/V$ (= [V(295K)-V(90K)]/V(295K). Note that $\Delta V/V$ in **(c)** is only for $90K \leq T \leq 295$ K, and this value is much greater for $90K \leq T \leq T_{MI} \sim 380$ K.

**Fig.2.** For $Ca_2Ru_{1-x}Mn_xO_4$ with x = 0.10, temperature dependences of **(a)** lattice parameters **a**-, **b**- and **c**-axis (right scale), **(b)** unit cell volume V, and **(c)** magnetic susceptibility $\chi_{ab}$ at $\mu_oH = 0.5$ T (field cooled) and ab-plane resistivity log $\rho_{ab}$ (right scale). The shaded area indicates the concomitant occurrence of the NTE and MI transition.

**Fig.3.** For $Ca_2Ru_{1-x}Mn_xO_4$ with x = 0.25, temperature dependences of **(a)** lattice parameters **a**-, **b**- and **c**-axis (right scale), **(b)** unit cell volume V, and **(c)** the magnetic susceptibility $\chi_{ab}$ at $\mu_oH = 0.05$ T and the ab-plane resistivity $\rho_{ab}$ (right scale).

**Fig.4.** For $Ca_2Ru_{1-x}Mn_xO_4$ with $0 \leq x \leq 0.25$, temperature dependence of **(a)** the ab-plane resistivity $\rho_{ab}$, **(b)** the specific heat C(T) and **(c)** magnetic suscepitibilty $\chi_{ab}$; **(d)** the isotheraml magnetization $M_{ab}$. Inset: enlarged C(T) near $T_N$. Note that $T_N$ and the order moment increase with x.

**Fig. 5.** For $Ca_2Ru_{1-x}Fe_xO_4$ with $0 \leq x \leq 0.22$, x dependence of **(a)** the lattice parameters **a**-, **b**- and **c**-axis (right scale) at T = 90 K, **(b)** unit cell volume V at T = 90 K and 295 K, and **(c)** thermal expansion ratio $\Delta V/V$ (= [V(295K)-V(90K)]/V(295K) for $0 \leq x \leq 0.22$

**Fig.6.** For $Ca_2Ru_{1-x}$**Fe**$_xO_4$ with x = 0.08, temperature dependences of **(a)** lattice parameters **a**-, **b**- and **c**-axis (right scale), **(b)** lattice volume V and specific heat C(T) (right scale) and **(c)** magnetic susceptibility $\chi_{ab}$ at $\mu_oH$ = 0.5 T (field cooled) for both x = 0.08 and 0.12, as indicated, and ab-plane resistivity log $\rho_{ab}$ (right scale). Note the unusual magnetization reversal for x = 0.12.

**Fig.7.** Temperature dependences of: **(a)** lattice parameters **a**-axis for x = 0, 0.10 (Mn) and 0.08 (Fe); **(b)** the Ru/M-Ru/M bond length **d** and the Ru/M-O1-Ru/M bond angle $\theta$ (right scale) for x = 0.08 (Fe); and **(c)** schematics illustrating changes of **d** and $\theta$ on cooling.

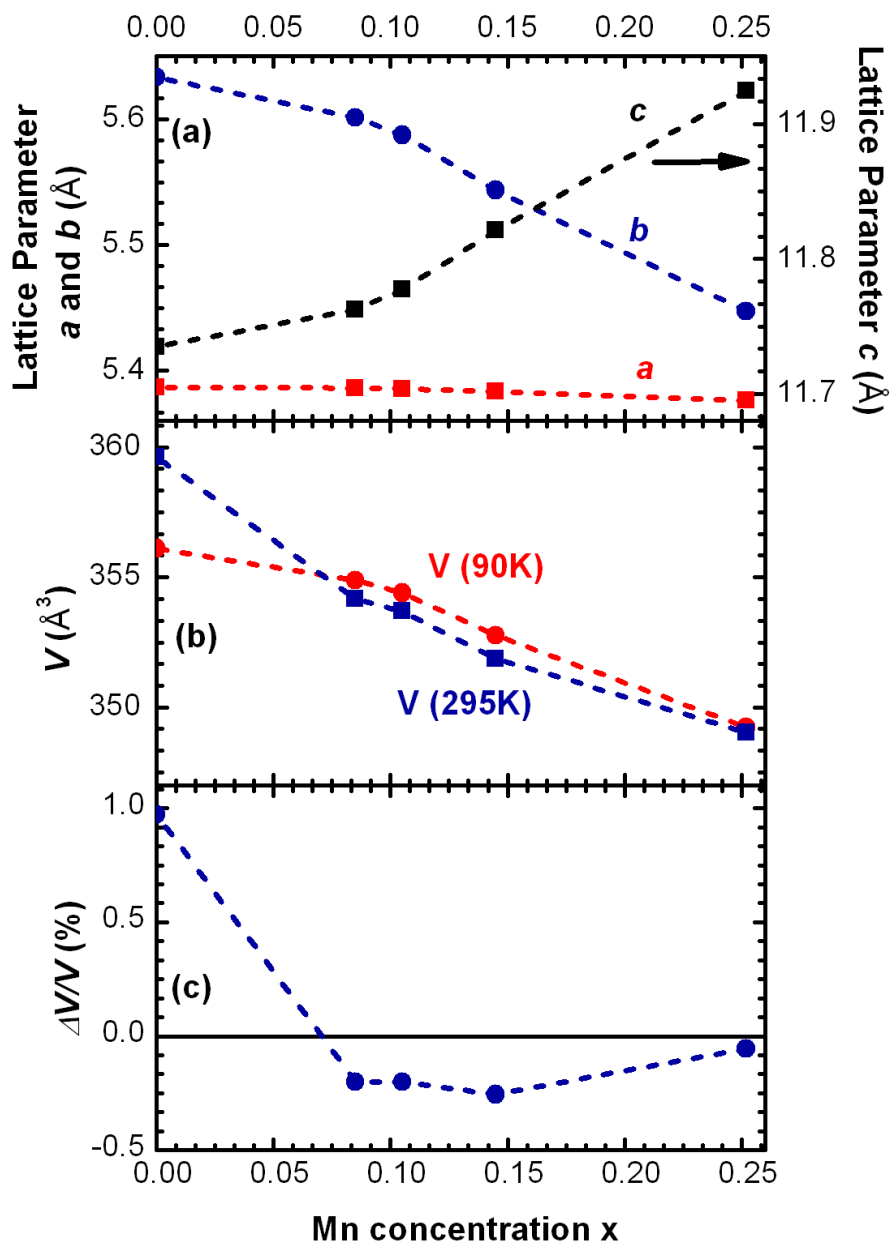

Fig.1

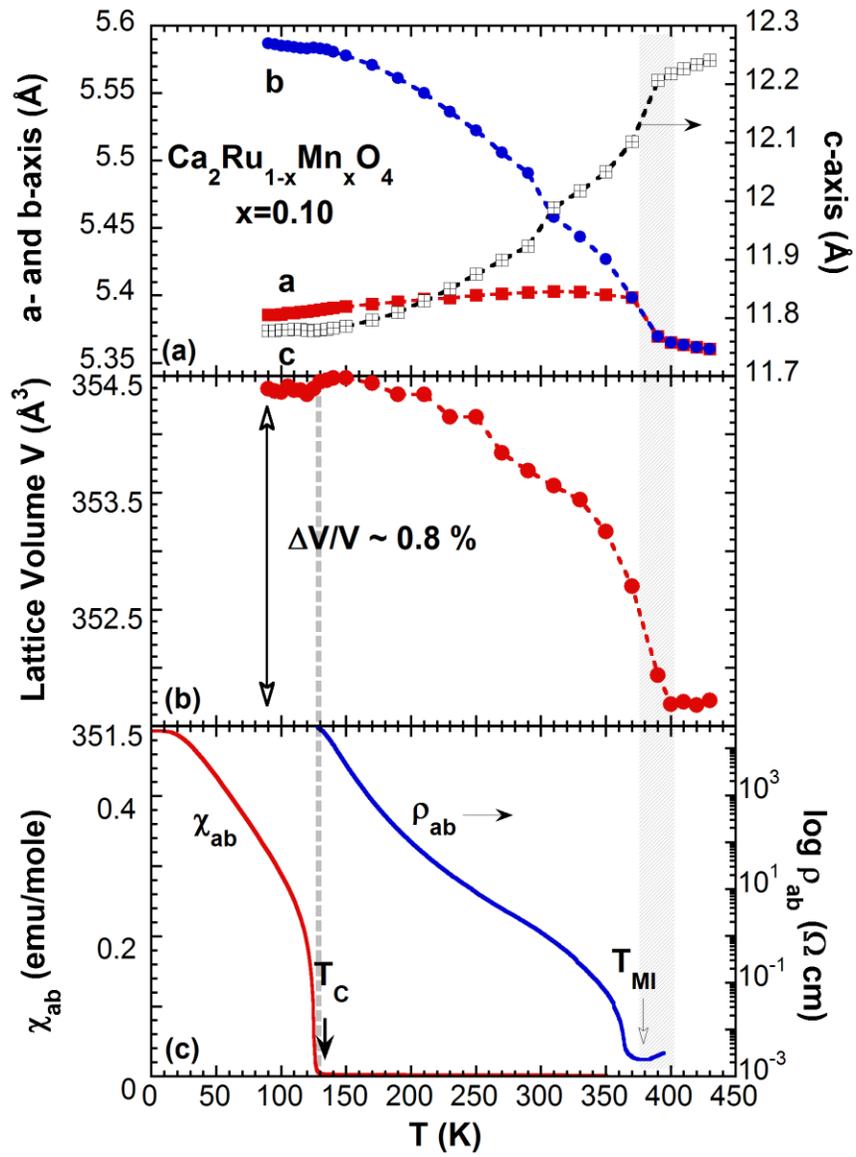

Fig.2

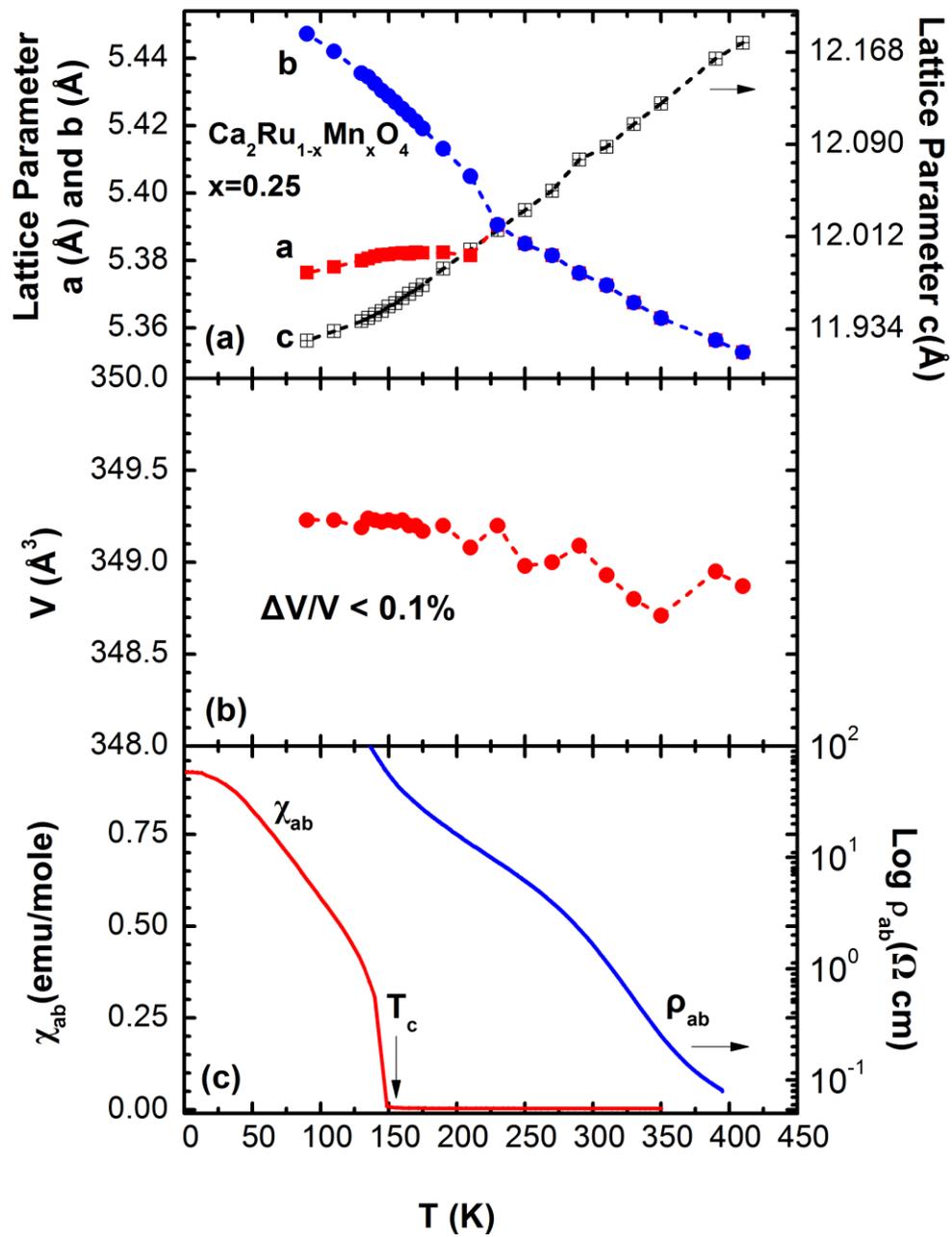

Fig. 3

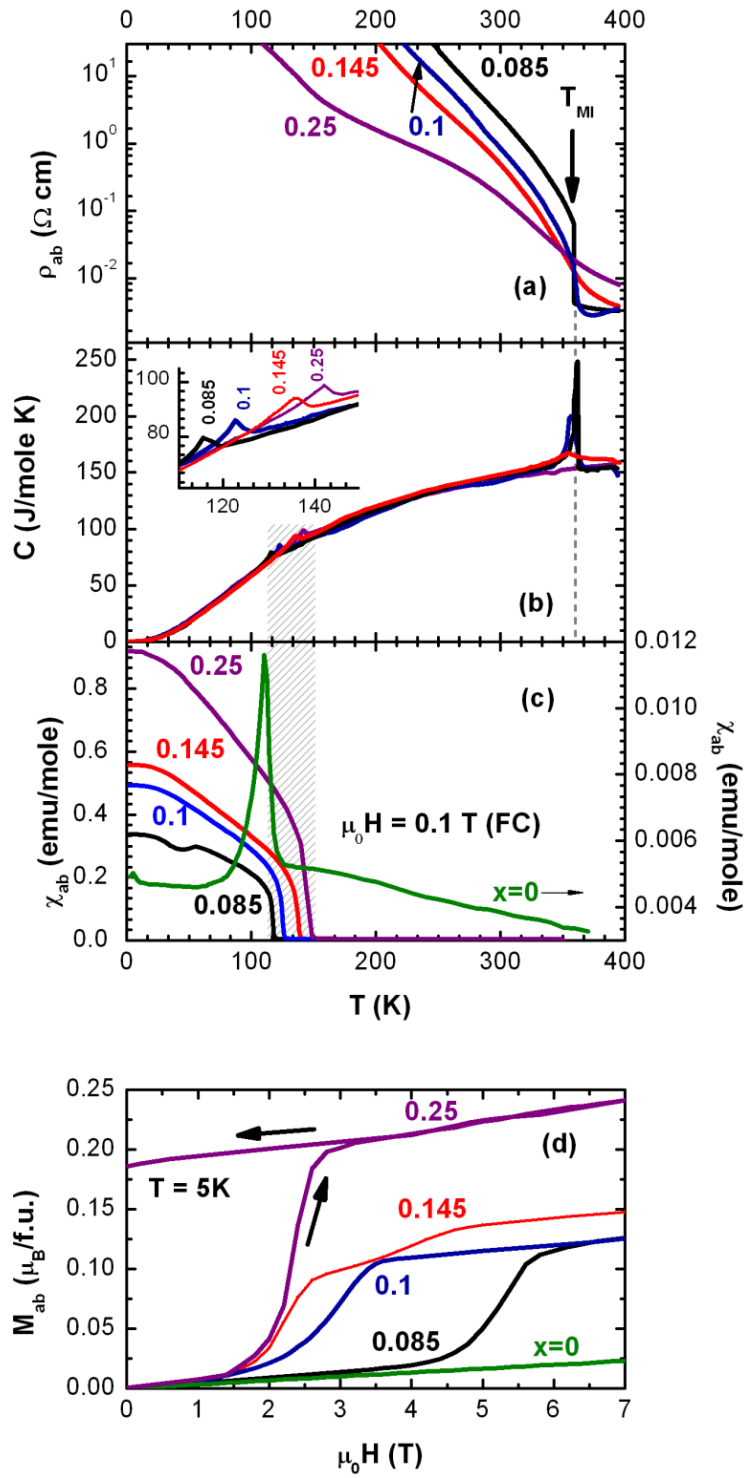

Fig. 4

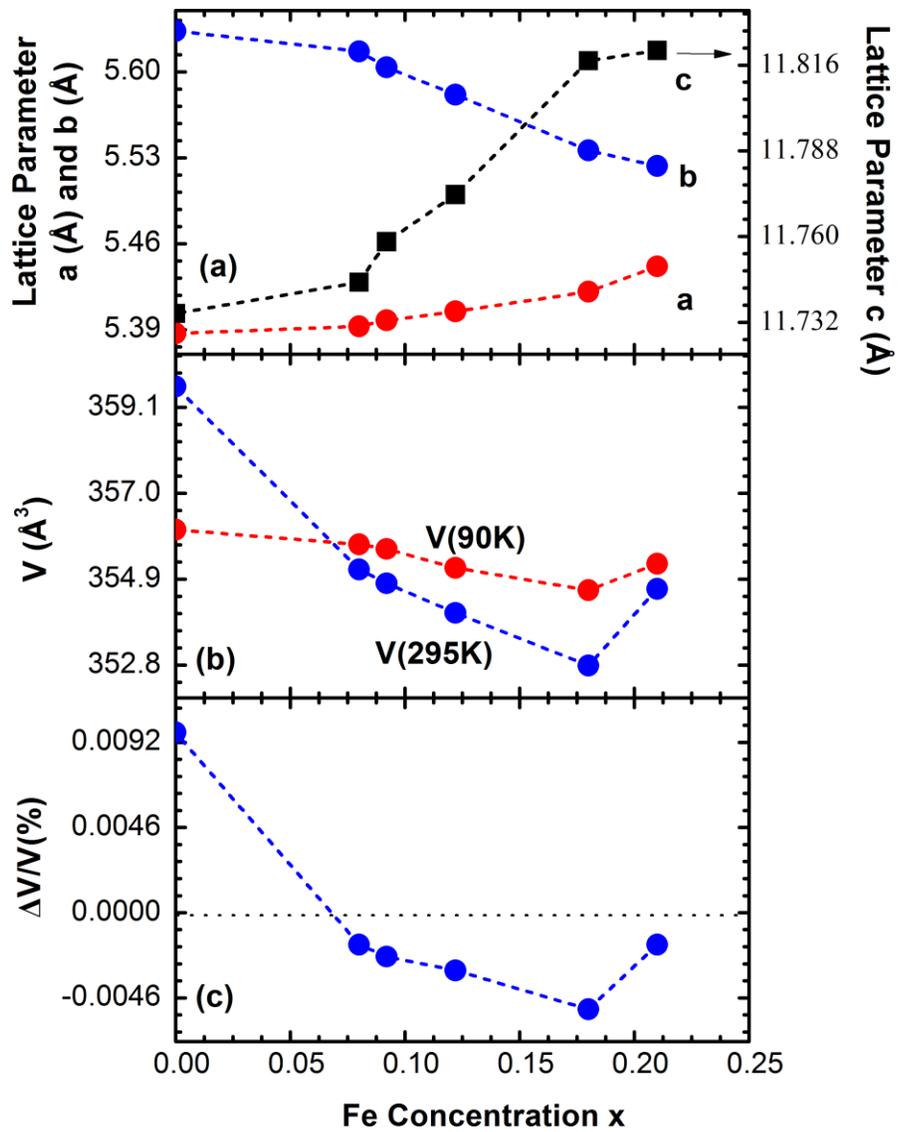

Fig.5

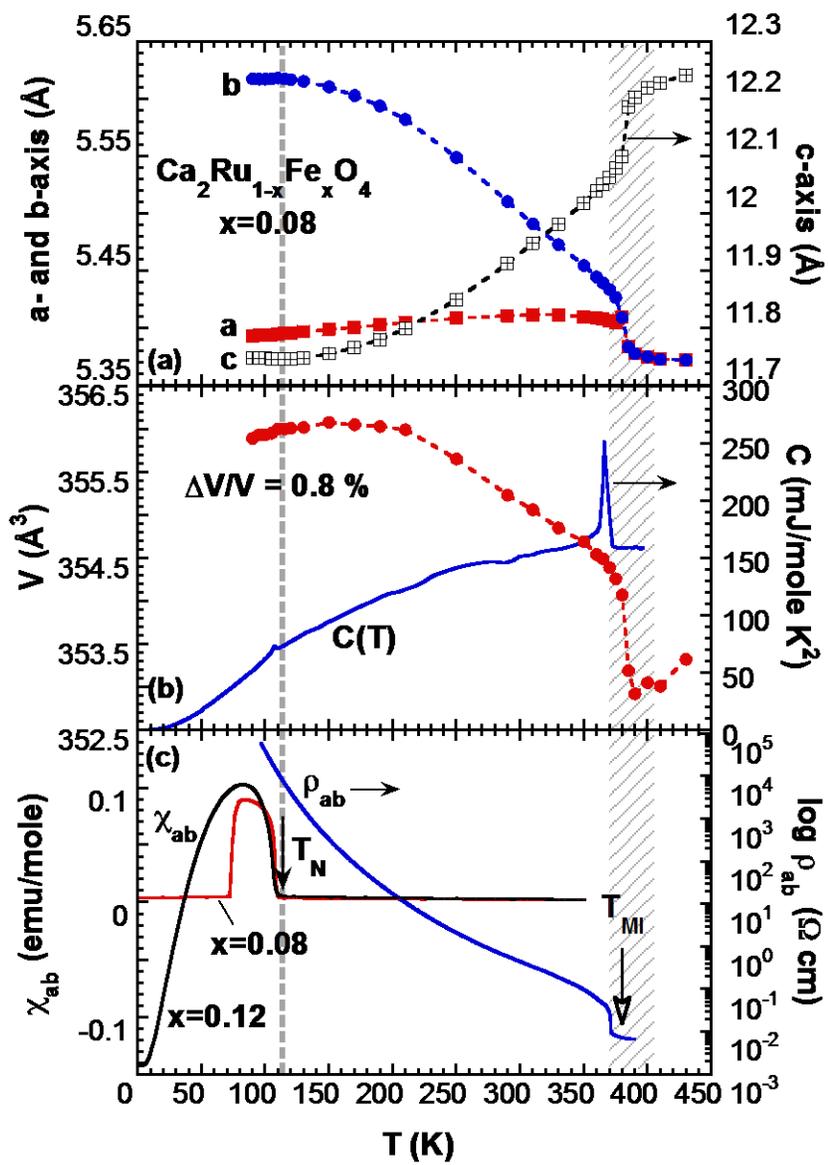

Fig.6

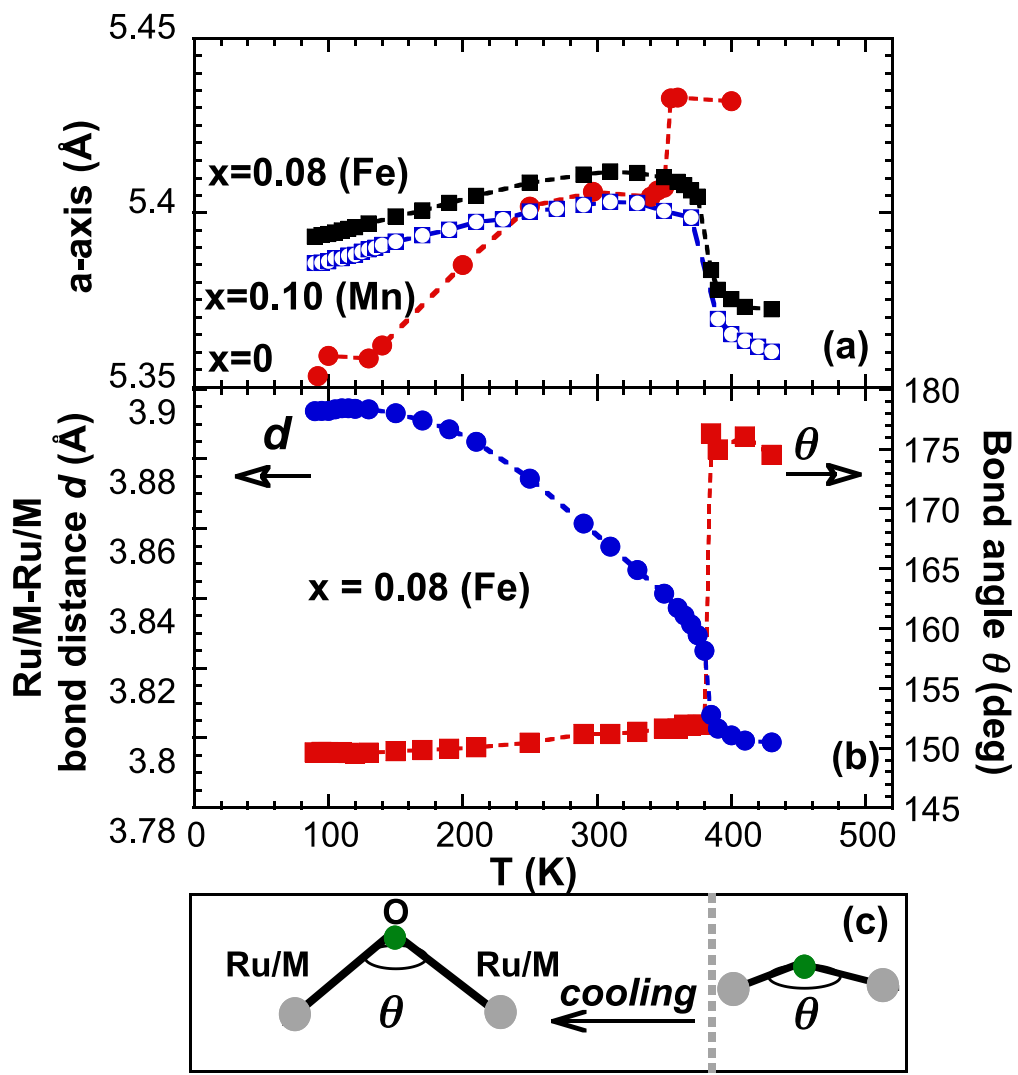